\documentclass[aps,prl,letterpaper,11pt,twoside,tightenlines,nofootinbib,showpacs,preprint,twocolumn]{revtex4}
\usepackage{graphicx}
\usepackage[sort&compress]{natbib}
\usepackage{latexsym}
\usepackage{epsfig}
\begin{document}
\newcommand{\eg}{{\it e.g.}}
\newcommand{\etal}{{\it et. al.}}
\newcommand{\ie}{{\it i.e.}}
\newcommand{\be}{\begin{equation}}
\newcommand{\ee}{\end{equation}}
\newcommand{\bea}{\begin{eqnarray}}
\newcommand{\eea}{\end{eqnarray}}
\newcommand{\bef}{\begin{figure}}
\newcommand{\eef}{\end{figure}}
\newcommand{\bce}{\begin{center}}
\newcommand{\ece}{\end{center}}
\def\lsim{\mathrel{\rlap{\lower4pt\hbox{\hskip1pt$\sim$}}
    \raise1pt\hbox{$<$}}}         
\def\gsim{\mathrel{\rlap{\lower4pt\hbox{\hskip1pt$\sim$}}
    \raise1pt\hbox{$>$}}}         

\title{Noncommutativity and Lorentz Violation in Relativistic Heavy Ion Collisions}
\author{P.~Castorina${}^{1,2}$, A.~Iorio${}^3$, D.~Zappal\`a${}^2$}

\affiliation{
\mbox{${}^1$ Dipartimento di Fisica, Universit\`a di Catania, Via Santa Sofia 64,
I-95123 Catania, Italy.}\\
\mbox{${}^2$ INFN, Sezione di Catania, I-95123 Catania, Italy.}\\
\mbox{${}^3$ Faculty of Mathematics and Physics, Charles University in Prague}\\
\mbox{V Hole\v{s}ovi\v{c}k\'ach 2, 18000 Prague 8 - Czech Republic}
}
\date{\today}
\begin{abstract}
\vskip10pt

The experimental detection of the effects of noncommuting coordinates in electrodynamic phenomena depends on the
magnitude of $|\theta B|$, where $\theta$ is the noncommutativity parameter and $B$ a background magnetic field. With
the present upper bound on $\theta$, given by $\theta_{\rm bound} \simeq 1/(10 \phantom{..} {\rm TeV})^2$, there was no
large enough magnetic field in nature, including those observed in magnetars, that could give visible effects or,
conversely, that could be used to further improve $\theta_{\rm bound}$. On the other hand, recently it has been
proposed that intense enough magnetic fields should be produced at the beginning of relativistic heavy ion collisions.
We discuss here lepton pair production by \textit{free} photons as one kind of signature of noncommutativity and
Lorentz violation that could occur at RHIC or LHC. This allows us to obtain a more stringent bound on $\theta$, given
by $10^{-3} \; \theta_{\rm bound}$, if such ``exotic'' events do not occur.

\end{abstract}
 \maketitle

High energy heavy-ion collisions are the most important tool to investigate the behavior of quantum chromodynamics
(QCD) at large temperature and density. The analysis of QCD phase diagram has required a great theoretical effort and the initial understanding in terms of a phase transition from confined quarks and gluons to a weakly interacting quark-gluon plasma (QGP) has been deeply modified by the experimental results at the Relativistic Heavy Ion Collider (RHIC). The new scenario of a strongly interacting QGP will be definitively clarified by future experiments at the Large Hadron Collider (LHC).

On the other hand, it has been recently shown \cite{kharzeev1,kharzeev2,kharzeev3} that other fundamental aspects of QCD, related to the topological nature of its vacuum, can be directly tested by experimental observations in relativistic heavy ion collisions. Indeed, gluon
field configurations with non zero topological charge generate chiral asymmetry, inducing $P$ and $CP$ violating
effects which produce an asymmetry between the amount of positive/negative charge above and below the reaction plane
\cite{kharzeev1,kharzeev2,kharzeev3}.

STAR Collaboration presented  \cite{star1} the conclusive observation of charge-dependent azimuthal correlations,
however the explanation of this charge asymmetry by $P-$ and $CP$-odd dynamics requires that a strong magnetic field is
produced at the beginning of the collision \cite{nota}. Analytical calculations \cite{kharzeev2} and numerical simulations
\cite{skolov} show that it is possible to produce an extremely intense magnetic field $|B| \simeq m_\pi^2$ in peripheral
heavy ion collisions at RHIC. What we want to stress here is that the production of magnetic fields of such intensity could
open a window on the detection of fundamental properties of spacetime through the manifestation of effects of the
noncommutativity of coordinates \cite{nc} and the associated violation of Lorentz symmetry \cite{chk}.

The literature on the various mechanisms for Lorentz violation is vast and includes the Standard Model Extension
\cite{coll}, theories with speed of light differing from $c$ and various other string theory/quantum gravity-inspired
effective field theories \cite{li}. In all cases, though, the bounds on the Lorentz violating parameters make the
experimental appreciation extremely elusive. For example, a low energy remnant of quantum gravity is posited to be a
modification of the dispersion relation for the photon given by, ($c=1$),
\be\label{uno}
E ^2 - k^2 = -  \xi k^3 / M_{Pl}
\ee
where $M_{Pl} \simeq 10^{19}$~GeV is the Planck mass and $\xi$ is a parameter that should be smaller than $10^{-15}$ 
\cite{jacobson}.

Let us now turn our attention to Maxwell theory in a noncommutative spacetime where $x^\mu \star x^\nu - x^\nu \star x^
\mu = i \theta^{\mu \nu} \;$, with $\star$ the Moyal-Weyl product, $\theta^{\mu \nu}$ an antisymmetric constant tensor 
(see, e.g., \cite{wess}) and $\mu, \nu = 0,1,2,3$. The general recipe to deal with gauge theories in such spacetimes 
was given in \cite{sw}. We are interested here on the $O(\theta)$-corrected action for Maxwell theory given by
\begin{eqnarray}
\hat{I} & = & - \frac{1}{4} \int d^4 x \; [F^{\mu \nu} F_{\mu \nu} \nonumber \\
& - & \frac{1}{2} \theta^{\alpha \beta} F_{\alpha \beta} F^{\mu \nu}
F_{\mu \nu} + 2 \theta^{\alpha \beta} F_{\alpha \mu} F_{\beta \nu} F^{\mu \nu}], \label{othetamaxwell}
\end{eqnarray}
with $F_{\mu \nu} = \partial_{\mu} A_\nu - \partial_{\nu} A_\mu$, and $A_\mu$ the \textit{usual} Abelian gauge field.
Clearly, noncommutative electrodynamics (NCED) is essentially a nonlinear generalization of Maxwell electrodynamics and
it was shown \cite{jack} that plane waves exist and, while those propagating along the direction of a background
magnetic field $\vec{B}$ still travel at the usual speed of light, those which propagate transversely to $\vec{B}$
have a modified dispersion relation given by
\begin{equation}\label{due}
 \omega = k (1 - \vec{\theta}_T \cdot \vec{B}_T)
\end{equation}
where $\vec{\theta} \equiv (\theta^{1}, \theta^{2}, \theta^{3})$ is the spatial part of $\theta^{\mu \nu}$ ($\theta^i =
\frac{1}{2} \epsilon^{ijk} \theta^{jk}$, $i,j,k=1,2,3$, the temporal components are taken to be zero $\theta^{0i}=0$)
and the subscript $T$ indicates the transverse component with respect to $\vec k$. 
That the quantum theory of NCED is sound is still an open issue due to certain novel divergencies that might \cite
{minwallaseiberg} or might not \cite{balachandran} appear in such theories, depending on the mathematical framework one 
uses for noncommutativity. Nonetheless, many phenomenological implications can be studied within the classical frame or 
by focusing on the kinematical
bounds on quantum processes. With this approach NCED, where the breaking of Lorentz invariance is directly related to 
noncommutativity
\cite{ncnoether}, has been discussed by considering synchrotron radiation \cite{noi1}, \v{C}erenkov effect in vacuum 
\cite{noi2}, ultra
high energy gamma rays \cite{noi3}. In all these works it was concluded that the effects of a nonzero $\vec{\theta}$ in 
NCED are very hard to detect due to the actual upper bound $\theta_{\rm bound} \simeq 1/(10 {\rm TeV})^2$ \cite
{chk,bertolami} and to the unavailability in nature of intense enough magnetic fields. 

According to our previous discussion, in 
the initial state of relativistic heavy ion collisions, 
for large impact
parameters, the magnetic field at the center of the collision can be approximately
written as, \cite{kharzeev2,skolov,kharzeev4} 
\begin{equation}\label{eq:magfield}
B(t) = \frac{1}{\left[1+ (t/\tau)^2\right]^{3/2}} B_0,
\end{equation}
with $\tau = b / (2\sinh Y)$,  $B_0 = 8 Z \alpha_{\mathrm{EM}} \sinh Y / b^2$ where $b$ denotes the impact parameter,
$Z$ the charge of the nucleus, and $Y$ the beam rapidity. For Gold-Gold ($Z=79$) collisions at RHIC (at 100 GeV per
nucleon)  one has $Y=5.36$ and, at typical large impact parameters  $b=10\;\mathrm{fm}$,  one finds $B_0 \sim 1.9
\times 10^5 \;\mathrm{MeV^2}$ and $\tau = 0.05\; \mathrm{fm}/c$.

From here it appears that the initial magnetic field in relativistic heavy ion collisions is much larger than the
magnetic fields of magnetars and one can consider the possibility of detection of the Lorentz violating nonzero $\vec
{\theta}$ 
effects. With these extremely intense magnetic fields one gains for $|\vec \theta_T \cdot  
\vec B_T|$ more than 20 orders of
magnitude with respect to magnetic fields in terrestrial laboratories. 

Before the evaluation of the impact of such gain on the dispersion relations in Eq.
(\ref{due}), which will be our main point here, let us give first an estimate of the impact on the noncommutative 
corrections to the synchrotron radiation spectrum of an energetic quark
traveling in the magnetic field produced at the beginning of the collision.

The synchrotron radiation spectrum in NCED has been evaluated in \cite{noi1}. Let us call $\omega$  the radiation
frequency, $\omega_0 \sim 1 / |\vec{r}|$  the cyclotron frequency, $\omega_c = 3 \omega_0 \gamma^3$ the critical
frequency, where $\vec{r}$ is the radius of the orbit and $\gamma$  the Lorentz factor. In the range  
$1 << \omega / \omega_0 << \gamma^3$, the ratio of the $\theta$-corrected energy $I$ (radiated in the plane of the 
orbit at large distances from the origin, within an angle $d\Omega$) to the standard one $I_{\theta=0}$, 
 in the most favourable case  $\vec \theta_T || \vec B_T $, is given by
\begin{equation}\label{ratio}
X \equiv \frac{d I (\omega) / d \Omega}{d I(\omega) / d
\Omega|_{\theta = 0}} \sim 1 + 20 \left( \frac{\omega_0}{\omega} \right)^{2/3}
|\theta B| \gamma^4 \;.
\end{equation}
For light energetic valence quarks in the heavy ion beams of 100 GeV per nucleon  the Lorentz factor could easily be
$\gamma \simeq O(10^2-10^3)$ and by considering  $\omega_0 \simeq 1/\tau$, where $\tau$ has been introduced in 
Eq.(\ref{eq:magfield}),
the  range  $1 << \omega / \omega_0 << \gamma^3$ allows the production of high frequency radiation. With $|\theta B|
\simeq 10^{-9}$, for Gold-Gold collisions at RHIC,  the ratio $X$ can be larger than 1 indicating that
either the spectrum is the $\theta$-corrected or, conversely, that the $\theta_{\rm bound}$ used must be ameliorated.
We do not proceed further with this case because to go beyond a mere indication of the effect here we should consider
the situation in greater detail (higher order contributions, time-varying magnetic fields, parton distribution of the
initial nucleon momentum, etc.).

What we can instead reliably focus on here are the dispersion relations in Eq.(\ref{due}) regarded as the kinematical
threshold for event obviously forbidden in standard electrodynamics, i.e. the pair production from a {\it single free}
photon $\gamma  \rightarrow e^+ e^-$.  In Eq.(\ref{due}) the  noncommutative contribution depends on the angle between $
\vec B_T$ and $\vec \theta_T$. At first order in $\theta$ one has
\be\label{tre}
{\omega}( 1 + \vec \theta _T \cdot \vec B_T)  = k
\ee
or
\be\label{quattro}
E_\gamma ^2 - k^2 = - 2 E_\gamma ^2 ( \vec \theta _T \cdot \vec B_T).
\ee
where $E_{\gamma} \equiv \omega $ .

Defining by $p_\gamma, p_+$ and $p_-$  the four momenta of $\gamma$, of  $e^+$ and of $e^-$,
respectively, the kinematical condition for the decay of a free photon whose action is given by (\ref{othetamaxwell}) is
\be\label{otto}
 - 2 E_\gamma ^2 ( \vec \theta _T \cdot \vec B_T) =( p_+ + p_-)^2 > 4 m_e^2
\ee
which requires  $(\vec \theta _T \cdot \vec B_T) < 0$. Therefore if in a heavy ion collision at RHIC one produces a
magnetic field of about $10^5$~MeV$^2$, with $\theta \simeq \theta_{\rm bound}$, a free photon can produce an $e^+
e^-$ pair. Hence an enhancement of lepton pairs with low invariant mass could be the signal of the ``exotic'' effects
of noncommutativity. With a magnetic field $B \sim 2\times 10^5 \;\mathrm{MeV^2}$ a photon with energy $E_\gamma = 50
\mathrm{GeV}$ opens a new channel for lepton pairs of invariant mass of about $2.5 \mathrm{MeV}$, while for $B \sim 2
\times 10^6 \;\mathrm{MeV^2}$ and $E_\gamma =100 \mathrm{GeV}$  the invariant mass is about $7.5 \mathrm{MeV}$.

Only $\gamma$s produced at the very beginning of the collisions have to be considered because the magnetic field
rapidly decreases with time (see Eq.(\ref{eq:magfield})) and noncommutative effects, if any, fade away accordingly.
Those high energy $\gamma$s should produce pairs of very small invariant mass (see Eq.(\ref{otto})). 
This implies that the electron
and positron are essentially produced collinearly but, since such ``exotic'' decay occurs at the very beginning of the
collision, one should observe the $e^+$ and $e^-$ 
with very large energy and an initial opposite curvature. Thus this pair production mechanism is completely different 
from: (a) the standard Drell-Yan process, which applies for large invariant masses;
(b) resonances decay, because one is 
considering $e^+ e^-$ with invariant mass much smaller
than the resonance masses; (c) the usual $\gamma \gamma  \rightarrow e^+ e^-$ and/or pair production in inhomogeneous
electromagnetic field. The latter pairs, however, could produce a large background.

The first step to single out the effects of the magnetic field is to consider the ratio between the yields of low
invariant mass lepton pairs in nucleus-nucleus (A-B) and proton-proton (p-p) collisions because in the latter the 
collective magnetic field is negligible \cite{kharzeev2}. However in heavy ion collisions rescattering effects still  
generate a significant background and therefore
the detection of the $\gamma$ decay allowed by noncommutative spacetime requires a more careful analysis.

Let us define the laboratory frame, say $(\hat{x}, \hat{y}, \hat{z})$, in such a way that the  reaction plane 
corresponds to the $\hat{x}-\hat{y}$
plane so that the magnetic field $B$ is produced in the $\hat z$ direction \cite{kharzeev2}. The noncommutative effect 
are enhanced
if the magnetic field transverse to the direction of the momentum of the photon  is maximum (see Eq.(\ref{otto}))
and then one has to focus on the reaction plane. In particular, if $\hat y$ is the beam axis,
to avoid the background in the forward direction, it is convenient to consider the $\gamma$ production in the reaction 
plane and with large
transverse momentum. For instance, if $\vec k\simeq (k_x , 0,0 )$ the noncommutative effect depend on the product
$\theta_z B$.

From this point of view a clear signal would be an enhancement, event by event, of  pairs of low invariant mass
in the reaction plane with respect to  pairs produced outside the reaction plane.

Moreover, the noncommutative parameter $\vec \theta$ is fixed in a non-rotating frame,
denoted by $(\hat X, \hat Y, \hat Z)$, whereas
 the component $\theta_z$ used above is defined in the previously introduced frame. Since this frame,
at fixed $\vec b$,
rotates with the earth, this component changes in time with the periodicity that depends on the earth's
sidereal rotation frequency $\Omega$.
By following the choice in \cite{bound2,chk}, one can take
the $\hat Z$ direction of the non-rotating frame coincident with the rotation axis
of the earth and $\hat X $ and $\hat Y$ with specific fixed celestial equatorial coordinates.
Then, by indicating with $(\theta_X,\theta_Y,\theta_Z)$ the components of the noncommutative parameter
in the non-rotating frame, one gets the explicit time dependence of $\theta_z$ \cite{bound2}
\bea\label{rotation}
\theta_z&=&(\sin \chi \cos \Omega t)\; \theta_X +
\nonumber\\
&&
(\sin \chi \sin \Omega t) \; \theta_Y + \cos \chi \; \theta_Z
\eea
where $\chi$ is the non-vanishing time-independent angle between the two axes $\hat Z$ and $\hat z$.
The oscillation in $\theta_z$ disappears in the peculiar case of $\vec \theta$ coincident with the
earth rotation axis (i.e. $\vec \theta=\theta_Z$ and therefore $\theta_z=\cos \chi \; |\vec \theta| $)
whereas it is maximal if $\vec \theta$ lies in the equatorial plane.
Apart from the unlikely case $\vec \theta=\theta_Z$,
Eq. (\ref{rotation}) clearly shows the oscillating structure of the product $( \vec \theta _T \cdot \vec B_T)$
which appears in
in Eq. (\ref{otto}) and  which, for the photons considered above, reduces to  the product $\theta_z B$.

Then a nonvanishing $\vec \theta$  induces a periodicity with frequency $\Omega$ in the number of
pairs produced at fixed $\vec b$.  In particular, for sufficiently high values of $B$ and $\theta$
to fulfill the bound in Eq. (\ref{otto}), by
looking at the ratio between pairs produced in (A-B)  and in (p-p) collisions
and by selecting those pairs generated by photons  with high transverse momentum
in the reaction plane,  so to reduce the effect of the background,
one should also be able to observe the periodic time dependence of this ratio,
in accordance with Eq. (\ref{rotation}).

If, on the contrary, the ratio of pair production number in (A-B)  and (p-p) shows no time dependence,
one can nonetheless establish a new bound on the components of $\vec \theta$ associated with the
oscillating factors, i.e. $\theta_X$ and $\theta_Y$,
based solely on the kinematical features which characterize the decay of
free $\gamma$ into $ e^+ e^-$. At LHC for Gold-Gold collision the magnetic field
could easily reach the  intensity $B_0 \simeq 3.2 $ GeV$^2$.
If a free photon with an energy of $E_\gamma \simeq 100$ GeV travels in such magnetic field
produced at the beginning of the collision and the pair production {\it is not observed},
then Eq. (\ref{otto}) gives the following bound
\be\label{nove}
\theta_{X,Y} < \frac{2 m_e^2}{B E_\gamma^2} \simeq \frac{1}{10^5 ({\rm TeV})^2} \simeq 10^{-3} \; \theta_{\rm bound}
\ee
where $\theta_{\rm bound}$ is the known bound \cite{chk}.

Strictly speaking, Eq.(\ref{due}) requires a constant background and one has to assume
that pair production indeed happens for short time $t<\tau$ after the collision 
(see Eq.(\ref{eq:magfield})).

To conclude: if extremely intense magnetic fields are produced in relativistic heavy ion collisions they might either 
help to observe 
effects of the noncommutative nature spacetime or, if such effects are not found, to establish a more stringent bound 
on $\theta$. 
Of course, a more detailed analysis is required for a direct comparison with experiments, nonetheless the enhancement 
of lepton pair production at low invariant mass we are suggesting here is a good probe as it is directly related with 
this new physics.

{\bf Acknowledgements} The authors thank D.~Kharzeev for very useful discussions and suggestions.

\end{document}